# Effects of Mn and Ti doping on superconductivity and charge ordering in $Na_xCoO_2$ system

Y. G. Shi, H. X. Yang, X. Liu, W. J. Ma, C. J. Nie, W. W. Huang and J.Q. Li*

Beijing National Laboratory for Condensed Matter Physics, Institute of Physics, Chinese Academy of Sciences, Beijing 100080, China

The superconductivity in $Na_{0.3}Co_{1-x}M_xO_2 \cdot 1.3H_2O$ and the charge ordering in $Na_{0.5}Co_{1-x}M_xO_2$ have been investigated for M = Mn and Ti substituting for Co. We have first successfully synthesized the single-phase $Na_{0.7}Co_{1-x}M_xO_2$ (M= Mn and Ti) materials with $0 \leq x \leq 0.1$, then we obtained $Na_{0.5}Co_{1-x}M_xO_2$ ($0 \leq x \leq 0.1$, M = Mn and Ti) by Na deintercalation and $Na_{0.3}Co_{1-x}M_xO_2 \cdot 1.3H_2O$ ($0 \leq x \leq 0.1$, M = Mn and Ti) by an additional water intercalations. X-ray diffraction measurements revealed that all samples are single-phase materials, their lattice parameters depend systematically on the Ti and Mn contents. Measurements of physical properties indicate that the superconductivity in $Na_{0.3}Co_{1-x}M_xO_2 \cdot 1.3H_2O$ is suppressed evidently by Co-site doping and killed at x = 0.02 for Mn doping and x = 0.01 for Ti doping. Charge ordering and magnetic properties in $Na_{0.5}Co_{1-x}M_xO_2$ were also influenced by M-atom doping.

Author to whom correspondence should be addressed: ljq@ssc.iphy.ac.cn.

The discovery of $Na_{0.3}CoO_2 \cdot 1.3H_2O$ superconducting compound [1] has attracted much attention as it is a new system other than cuprates in which a doped Mott insulator becomes a superconductor by charge carrier doping. In order to obtain deeper insight into physical properties and superconducting mechanism in this new system, theoretical and experimental investigations have been carried out on the $Na_xCoO_2$, $Na_{0.3}CoO_2 \cdot 1.3H_2O$ and some other related materials [2, 3, 4, 5]. Recently, two distinct metallic states separated by a charge ordered state are suggested in a phase diagram for non-hydrated $Na_xCoO_2$ [6]. This specific charge ordering was suggested to be in connection with a major instability in the narrow conduction band of $CoO_2$ layer [7]. Element doping, as a effective way, have been often used to study the essential charge correlation, even the feature of superconducting pairing, in the layered cuprate superconducting systems [8], Yokoi et al have investigated the Ir and Ga doping for Co and examined the effect on superconducting critical temperature $Tc$ in $Na_{0.3}Co_{1-x}M_xO_2 \cdot 1.3H_2O$ ($0 \leq x \leq 0.025$, M = Ga and Ir) [9], and discussed with pair symmetry in this new superconducting system. Similar investigations have been done about the physical and the structural research on the parent compounds $Na_{0.7}Co_{1-x}M_xO_2$ [10, 11]. In present paper, we will mainly report on structural and physical properties in $Na_{0.3}Co_{1-x}M_xO_2 \cdot 1.3H_2O$ and $Na_{0.5}Co_{1-x}M_xO_2$ ($0 \leq x \leq 0.1$, M = Mn and Ti) materials. We found the either Mn atom or Ti doping on the Co site can evidently affect the superconducting properties in the water-intercalated $Na_{0.3}CoO_2 \cdot 1.3H_2O$ materials and the charge ordering in $Na_{0.5}CoO_2$.

Polycrystalline samples of $Na_{0.7}Co_{1-x}M_xO_2$ ($0 \leq x \leq 0.1$, M = Mn or Ti) were synthesized by solid-state reactions as reported in ref.12, A amount of $Na_2CO_3$, $Co_3O_4$, and $MnO_2$ ($TiO_2$) are mixed and sintered for 8 ~ 12 h in air at the temperatures ranging from 800 to 850 °C. Then the products were pressed into pellets and sintered for another 8 hours at the same temperature range. The $Na_{0.5}Co_{1-x}M_xO_2$ (M = Mn or Ti) compounds were prepared through a Na deintercalation [6]. Superconducting $Na_{0.3}Co_{1-y}M_yO_2 \cdot 1.3H_2O$ materials were synthesized also by means of the Na-deintercalation following by a water intercalation just as the synthesis of superconducting $Na_xCoO_2 \cdot 1.3H_2O$ materials [2]. X-ray diffraction (XRD) measurements on all samples were carried out with a diffractometer in the Bragg–Brentano geometry using Cu Ka radiation. Low-temperature magnetization measurements as a function of temperature were performed using a commercial Quantum Design SQUID. The resistivity (R) as a function of temperature was measured by a standard four-point probe technique. Silver paint was used to make electrical contacts to the ceramic samples.

We have successfully synthesized samples with nominal compositions of $Na_{0.7}Co_{1-x}M_xO_2$ ($0 \leq x \leq 0.1$, M = Mn and Ti), certain physical and structural properties in doping Mn-atom material has been characterized and reported in our previous paper [11]. Careful structural analysis suggests that the $Na_{0.7}Co_{1-x}Mn_xO_2$ materials and the $Na_{0.7}Co_{1-x}Ti_xO_2$ materials with $0 \leq x \leq 0.1$ can be considered as single phase samples without any clear diffraction peaks from other impurities in XRD patterns. In the higher doping range (x > 0.3), certain impurities as minor phases appear in the products. Figure 1 shows a series of the XRD spectra obtained from a series of $Na_{0.7}Co_{1-x}Ti_xO_2$ samples with x ranging from 0 to 0.1. All diffraction peaks in these spectra can be well indexed on the known hexagonal unit cell with space group of $P6_3/mmc$. Table 1 lists the structural parameters obtained from our XRD experiments for samples with nominal compositions of $Na_{0.7}Co_{1-x}M_xO_2$ ($0 \leq x \leq 0.1$, M = Mn and Ti). It is clear that, within these two series, the lattice parameters changes systematically with x. This fact demonstrates that the expected substitution at within $CoO_2$ sheets has been successfully carried out in this doping range. It is noted that in both series the lattice parameter within *a-b* plane remains nearly constant, and, on the other hand, the *c* axis parameter of the unit cell increases monotonously with the increase of M-atom doping.

Substation of transition metal elements for Co within the $CoO_2$ sheets is a significant way to understand the superconductivity and charge correlation in the layered superconducting system, as demonstrated in the investigations of the high *Tc* superconducting Cu-oxides [8]. In present study, materials with nominal composition of $Na_{0.3}Co_{1-x}M_xO_2 \cdot 1.3H_2O$ ($0 \leq x \leq 0.1$, M = Mn, Ti) have been successfully synthesized. Figure 2 shows the XRD data obtained from these materials at room temperature. Analyses of these XRD spectra suggest that the products were nearly single-phase and all diffraction peaks in these patterns can be well indexed by the hexagonal cell with the same space group of $P6_3/mmc$ as reported [1]. Table 2 lists the lattice parameters *a*, *c* and unit cell volumes for different Mn and Ti doping contents. It is remarkable that the both lattice parameters and cell volume change slightly irregular in comparison with the $Na_{0.7}Co_{1-x}M_xO_2$ materials. This property is believed to arise from possible difference of intercalated $H_2O$ content or Na content, they both could change faintly from one sample to another, especially for the hydrated $Na_{0.3}Co_{1-x}M_xO_2 \cdot 1.3H_2O$ materials. The structural examinations indicate that in general the lattice parameter of $Na_{0.3}Co_{1-x}Mn_xO_2 \cdot 1.3H_2O$ ($x \leq 0.1$) samples varies in a range from 2.816 Å to 2.826 Å within the *a-b* plane and from 19.633 Å to 19.788 Å for *c* lattice parameters; the lattice parameters of $Na_{0.3}Co_{1-x}Ti_xO_2 \cdot 1.3H_2O$ ($x \leq 0.1$) materials range from 2.821 Å to 2.839 Å within *a-b* plane and from 19.668 Å to 19.833 Å along *c* direction.

Measurement of magnetization on these samples reveals that both Mn atom and Ti atom doping can evidently influence the superconductivity and the low temperature physical properties. Figure 3a and b show the zero-field cooling DC magnetization data measured in a field of 20 Oe for $Na_{0.3}Co_{1-x}M_xO_2·1.3H_2O$ (0 ≤ x ≤ 0.1, M = Mn or Ti,) samples. The superconducting transition at about 4.3 K in $Na_{0.3}CoO_2·1.3H_2O$ sample decreases quickly with the doping of either Mn atom or Ti atom on the Co site. The superconducting transition temperature in sample of $Na_{0.3}Co_{1-x}Mn_xO_2·1.3H_2O$ with x = 0.01 appears at around 3.8 K as demonstrated by the magnetization data as shown in figure 3a. The superconducting diamagnetic signal appears in this material is much weaker than that in the $Na_{0.3}CoO_2·1.3H_2O$ material. It is remarkable that the magnetic anomalies shown as small humps/peaks can be clearly recognized in magnetizations. This alternation is likely in connection with the competition between the superconducting state and a low-temperature antiferromagnetic state. As a result, the superconductivity vanishes totally in the x = 0.02 sample in which a noticeable antiferromagnetic signal appears at about 4.6 K. It is also noted that, between x = 0.02, and x = 0.05, the $Na_{0.3}Co_{1-x}Mn_xO_2·1.3H_2O$ samples show up complex magnetic properties at the temperature below 4.5 K possibly arising from coexistences of several kinds of magnetic orders. The x = 0.1 sample has a simple paramagnetic ground state as illustrated in figure 3a.

In the $Na_{0.3}Co_{1-x}Ti_xO_2·1.3H_2O$ (x ≤ 0.1) samples, experimental investigations suggest that the superconductivity is only visible up to x = 0.005 with $T_c$ = 3.8 K and totally disappears in the samples with x ≥ 0.01 as demonstrated in figure 3b. In x = 0.01 sample there is a noticeable antiferromagnetism signal at about 3.4 K. With the increase of Ti doping, the low temperature properties change progressively to a paramagnetic state as what observed in the Mn-doped materials. The most striking phenomenon noted in our experiments on $Na_{0.3}Co_{1-x}M_xO_2 \cdot 1.3H_2O$ ($x \leq 0.1$, M = Mn, Ti) is the presence of a common transformation of the low temperature states, i.e. from the superconducting state ($x = 0$), via the antiferromagnetic state ($x = 0.01 \sim 0.02$), towards the paramagnetic state. It well known that the magnetic impurities, such as Mn atoms, could have large effects in usual s-wave superconductors, we therefore have also made certain attempts to perform the comparison of the values of ($dT_c/dx$) with the data reported previously for the $Na_{0.3}Co_{1-x}(Ir, Ga)_xO_2 \cdot 1.3H_2O$ materials [9]. The results indicate that the slopes $dT_c/dx$ in both Mn and Ti doping systems is larger than that of Ir-doped materials. Because it is extremely difficulty to accurately control the water and Na contents in the nonstoichiometric superconducting phase, we have to consider more factors in a further systematical analysis on the effects of Mn and Ti doping on the physical properties in present system.

In order to study the effects of Mn and Ti doping at the Co sites on charge ordering transition observed in $Na_{0.5}CoO_2$ material, we have synthesized a series materials with nominal composition of $Na_{0.5}Co_{1-x}M_xO_2$ ($0 \leq x \leq 0.1$, M = Mn or Ti). Figure 4 shows the XRD spectra of the samples used in present study. Table 3 lists the lattice parameters *a*, *c* and unit cell volume for all samples with different Mn or Ti contents. The structural data indicate that, in Mn-doped samples, the *c* axis parameter increases progressively with the increase of Mn content and the *a* axis parameter almost remains a constant; and, in the Ti-doped samples, both *a* and *c* axis parameters increase progressively with the increase of Ti doping.

Figure 5a and b show respectively the magnetic susceptibilities from two series of samples $Na_{0.5}Co_{1-x}M_xO_2$ (M = Mn or Ti), illustrating the low temperature magnetic properties in correlation with the charge order transition in $Na_{0.5}CoO_2$ materials. It is notable that the magnetic susceptibility of the x = 0 sample shows three clear transitions at 90 K, 53 K and 25 K, respectively (figure 5a). The most evident anomaly shown up as a downturn occurs at around 53K in magnetization. In $Na_{0.5}Co_{1-x}M_xO_2$ (M = Mn ) materials, it is observed that doping very little Mn atom (about 0.01) could evidently affect the charge-order transition. Actually the major transition at 53 K becomes hardly visible in the x = 0.01 sample. The x = 0.03 sample undergoes a clear paramagnetic transition without any magnetic anomalies(figure 5a). In Ti-doped $Na_{0.5}Co_{1-x}M_xO_2$ system(figure 5b), the x = 0.01 sample also shows two clear magnetic transitions at the temperatures of around 58 K and 22 K; the x = 0.05 sample shows a clear magnetic transition at about 66 K; the x = 0.1 sample undergoes a clear paramagnetic transition without any magnetic anomalies in connection with charge ordering. Actually the major transition at 53 K becomes hardly visible in the x = 0.1 sample.

Figure 6 shows temperature dependent of resistivity for M doped samples. The shown resistances for all samples are normalized with the 300K resistance for facilitating the comparison. The experimental data for M-doped samples show clearly increase of resistances along with the rise of doping level. Figure 6a shows temperature dependent of resistivity for Mn doping samples with x = 0, 0.01, and 0.03. The transition is seen at 52 K in consistent with the results of magnetization with x = 0.01. And as doping Mn content above x = 0.03, the charge-order transition is invisible. Figure 6b demonstrates that all $Na_{0.5}Co_{1-x}Ti_xO_2$ samples with $0 \leq x \leq 0.05$ show the semiconducting behavior between 5 and 300 K. Charge ordering transition can only be recognized by dR/dT as illustrated in the inset of figure 6b. These results directly suggest that the magnetic doping (Mn) has a larger influence on the charge ordering. Both magnetization and resistivity are evidently changed with slight doping.

In summary, we successfully synthesized $Na_{0.5}Co_{1-x}M_xO_2$ and $Na_{0.3}Co_{1-x}M_xO_2 \cdot 1.3H_2O$ ($0 \leq x \leq 0.1$, M = Mn and Ti) materials. Their structure and low-temperature physical properties have been systematically investigated. XRD measurements suggest that the lattice parameters in all systems change systematically with x. Doping Mn or Ti atoms evidently affects the superconducting property as observed in $Na_{0.3}CoO_2 \cdot 1.3H_2O$ and the charge ordering in $Na_{0.5}CoO_2$. The superconductivity is totally destroyed in $Na_{0.3}Co_{1-x}M_xO_2 \cdot 1.3H_2O$ materials with $x > 0.02$ for Mn doping and $x > 0.01$ for Ti doping. Magnetic Mn doping in the charge ordering of $Na_{0.5}CoO_2$ material has evident stronger effects on both charge ordering and the magnetic properties than Ti doping.

**Acknowledgments**

We would like to thank Prof. Y.Q. Zhou, Prof. X.F. Duan and Miss Y. Li for their assistance in measurements of physical properties. The work reported here is supported by the National Nature Science Foundation of China.

Figure captions

Table 1. Lattice parameters a, c and unit cell volume of $Na_{0.7}Co_{1-x}M_xO_2$ (M = Mn or Ti).

Table 2. Lattice parameters a, c and unit cell volume of $Na_{0.3}Co_{1-x}M_xO_2\cdot1.3H_2O$ (M = Mn or Ti)

Table 3. Lattice parameter a, c and unit cell volume of $Na_{0.5}Co_{1-x}M_xO_2$ (M = Mn or Ti).

Figure 1. X-ray diffraction patterns of $Na_{0.7}Co_{1-x}Ti_xO2$.

Figure 2. XRD patterns for $Na_{0.3}Co_{1-x}M_xO_2\cdot1.3H_2O$, (M = Mn or Ti).

Figure 3. Temperature dependence of magnetization for $Na_{0.3}Co_{1-x}M_xO_2\cdot1.3H_2O$ samples, revealing the complex alternation of magnetic properties with Co-site doping. (a) M = Mn and (b) M = Ti.

Figure 4. X-ray diffraction patterns of $Na_{0.5}Co_{1-x}M_xO_2$ (M = Mn or Ti) samples.

Figure 5. The magnetic susceptibility $\chi$ as a function of temperature for $Na_{0.5}Co_{1-x}M_xO_2$ (M = Mn or Ti) samples, illustrating the changes of charge ordering in the doped samples.

Figure 6. Resistivity as a function of temperature for $Na_{0.5}Co_{1-x}M_xO_2$ (M = Mn or Ti) samples, demonstrating that magnetic Mn-doping has much more considerable effect on charge ordering. (a) M = Mn and (b) M = Ti.

**Table 1**

| | Doping content x | $a$(Å) | $c$(Å) | Unit cell volume(Å³) |
|---|---|---|---|---|
| $Na_{0.75}Co_{1-x}Mn_xO_2$ | 0 | 2.839 | 10.805 | 226.253 |
| | 0.01 | 2.832 | 10.890 | 226.910 |
| | 0.03 | 2.831 | 10.907 | 227.104 |
| | 0.05 | 2.832 | 10.912 | 227.368 |
| | 0.1 | 2.829 | 10.944 | 227.552 |
| $Na_{0.75}Co_{1-x}Ti_xO_2$ | 0 | 2.839 | 10.805 | 226.253 |
| | 0.01 | 2.835 | 10.887 | 227.328 |
| | 0.03 | 2.835 | 10.901 | 227.621 |
| | 0.05 | 2.833 | 10.924 | 227.779 |
| | 0.1 | 2.830 | 10.964 | 228.129 |

**Table 2**

|  | Doping content x | $a$(Å) | $c$(Å) | $c/a$ | Unit cell volume(Å$^3$) |
|---|---|---|---|---|---|
| $Na_y Co_{1-x}Mn_xO_2 \cdot H_2O$ | 0 | 2.825 | 19.788 | 7.005 | 410.278 |
|  | 0.01 | 2.823 | 19.693 | 6.976 | 407.730 |
|  | 0.02 | 2.816 | 19.633 | 6.972 | 404.474 |
|  | 0.03 | 2.823 | 19.671 | 6.968 | 407.275 |
|  | 0.05 | 2.824 | 19.703 | 6.977 | 408.226 |
|  | 0.1 | 2.826 | 19.747 | 6.988 | 409.718 |
| $Na_yCo_{1-x}Ti_xO_2 \cdot H_2O$ | 0 | 2.825 | 19.788 | 7.005 | 410.278 |
|  | 0.005 | 2.821 | 19.811 | 7.023 | 409.593 |
|  | 0.01 | 2.823 | 19.668 | 6.967 | 407.213 |
|  | 0.03 | 2.833 | 19.753 | 6.972 | 411.875 |
|  | 0.05 | 2.839 | 19.698 | 6.938 | 412.470 |
|  | 0.1 | 2.817 | 19.833 | 7.040 | 408.885 |

**Table 3**

|  | Doping content x | $a$(Å) | $c$(Å) | $c/a$ | Unit cell volume(Å$^3$) |
|---|---|---|---|---|---|
| $Na_{0.5}Co_{1-x}Mn_xO_2$ | 0 | 2.813 | 11.121 | 3.953 | 228.624 |
|  | 0.005 | 2.813 | 11.121 | 3.953 | 228.624 |
|  | 0.01 | 2.813 | 11.125 | 3.955 | 228.707 |
|  | 0.03 | 2.813 | 11.129 | 3.956 | 228.789 |
|  | 0.05 | 2.815 | 11.140 | 3.957 | 229.341 |
|  | 0.1 | 2.811 | 11.151 | 3.967 | 228.915 |
| $Na_{0.5}Co_{1-x}Ti_xO_2$ | 0 | 2.813 | 11.121 | 3.953 | 228.624 |
|  | 0.005 | 2.814 | 11.122 | 3.952 | 228.807 |
|  | 0.01 | 2.814 | 11.123 | 3.953 | 228.828 |
|  | 0.03 | 2.815 | 11.134 | 3.955 | 229.217 |
|  | 0.05 | 2.815 | 11.143 | 3.958 | 229.402 |
|  | 0.1 | 2.818 | 11.191 | 3.971 | 230.882 |

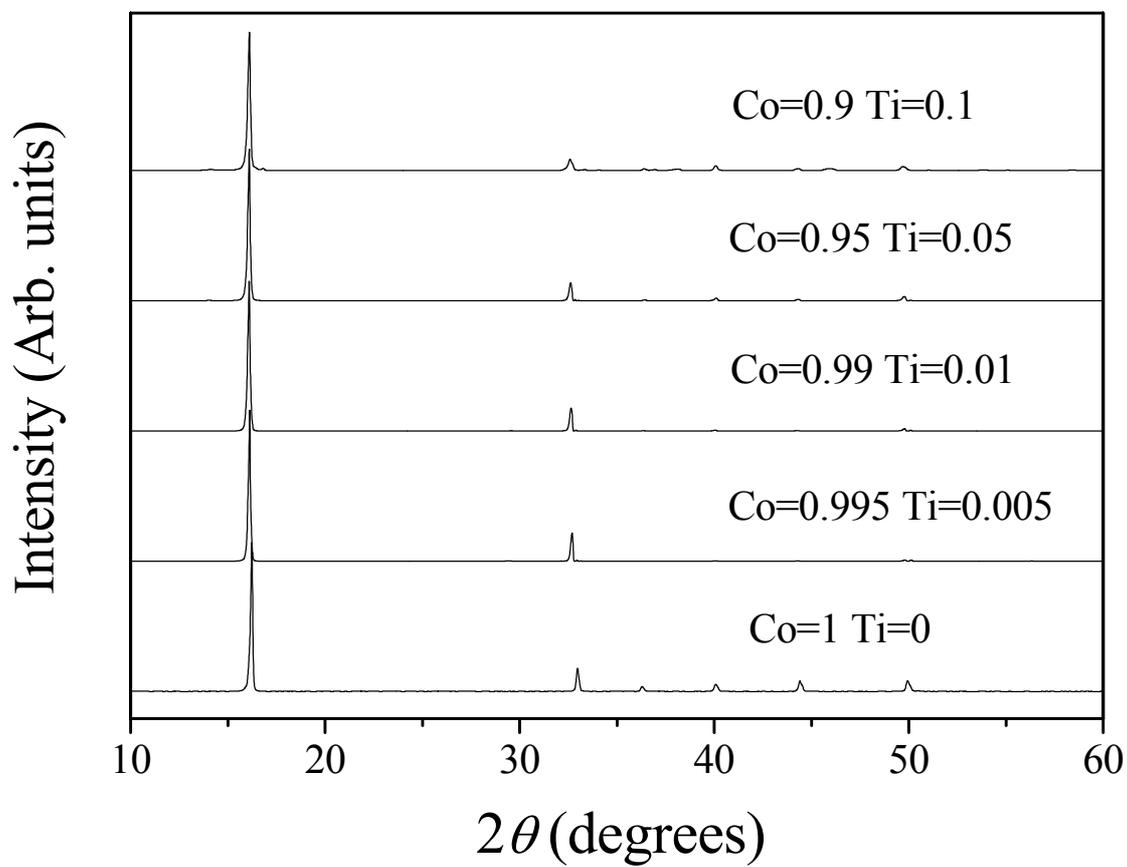

**Figure 1**

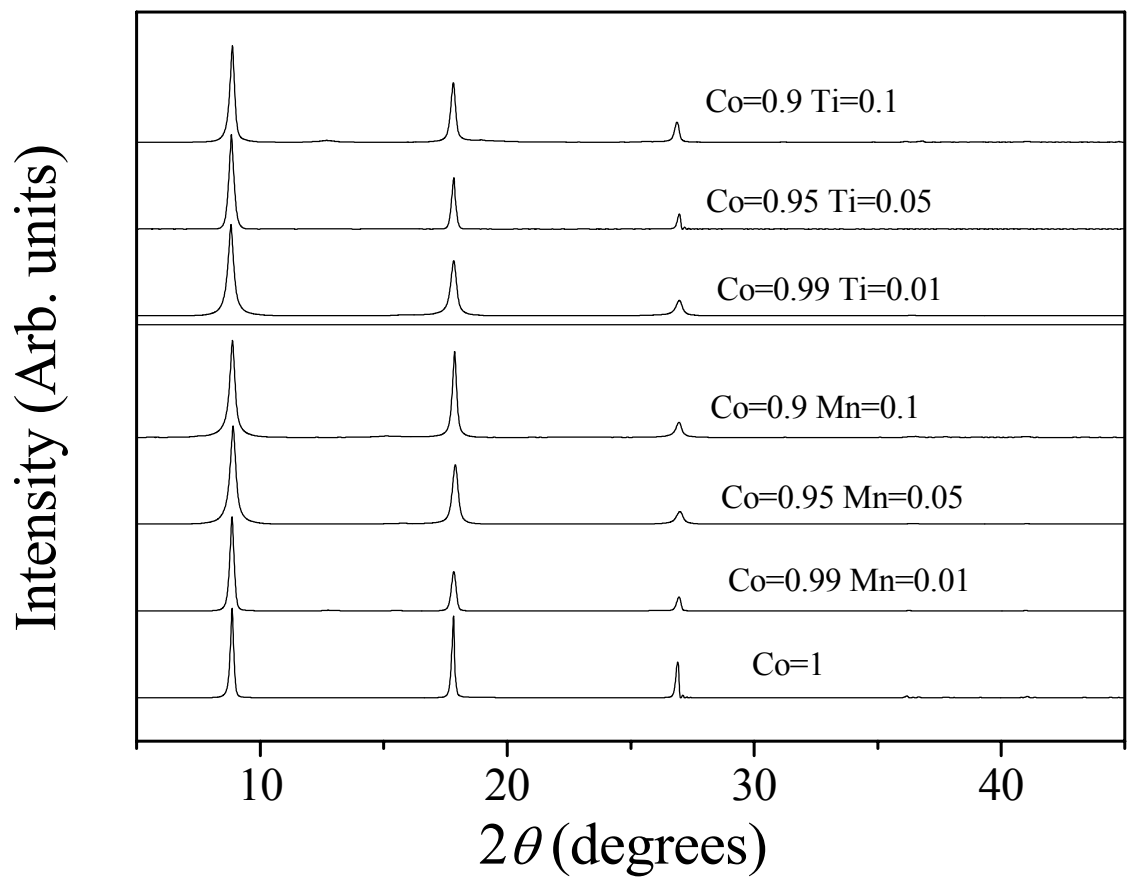

**Figure 2**

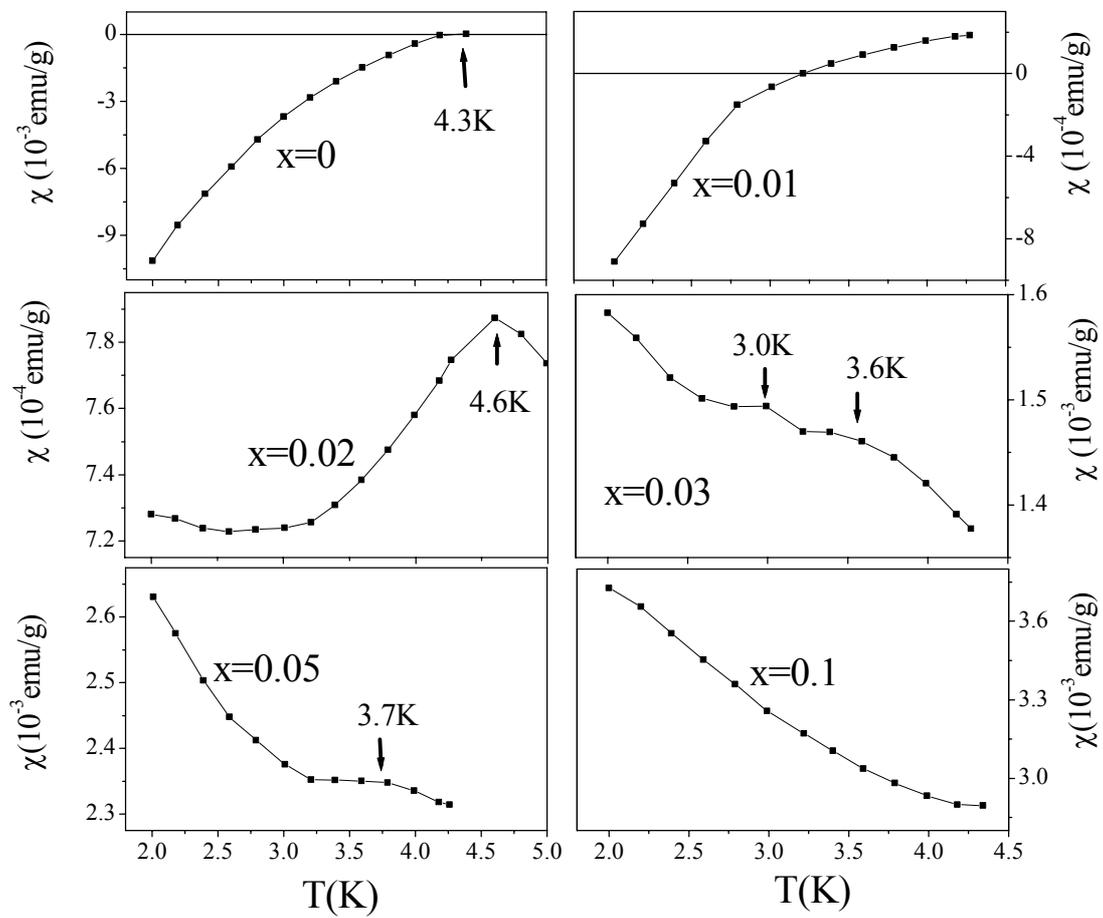

**Figure 3 (a)**

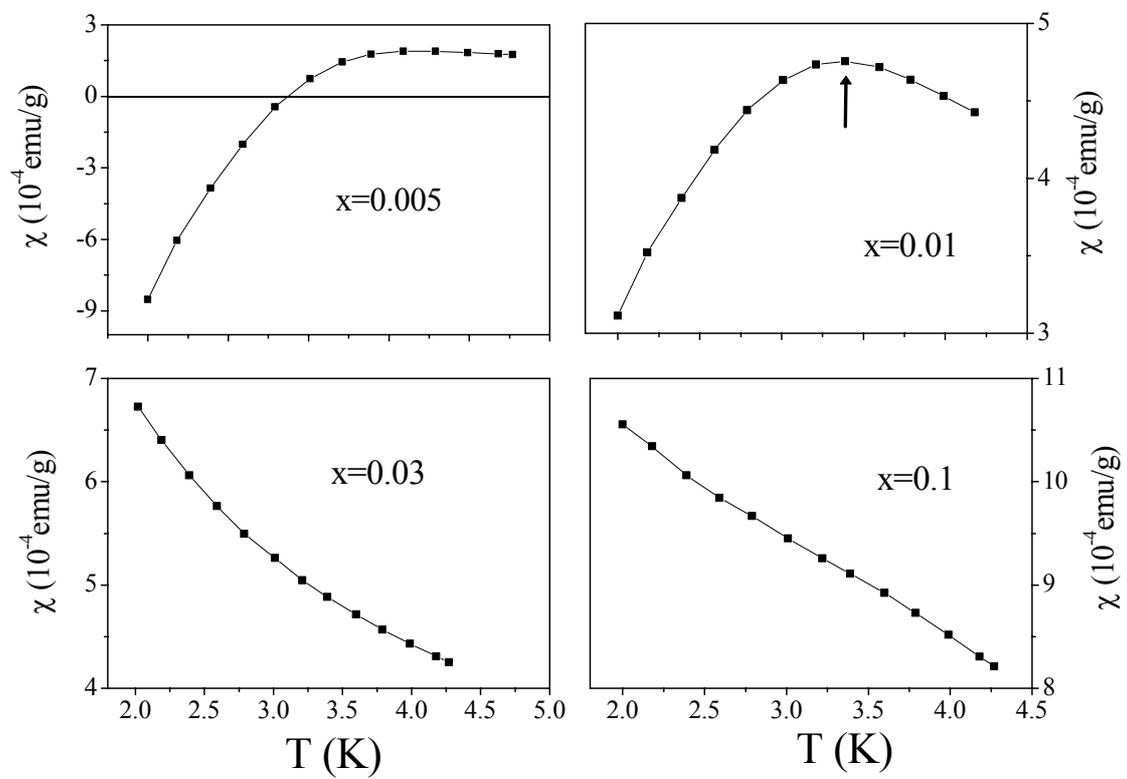

**Figure 3 (b)**

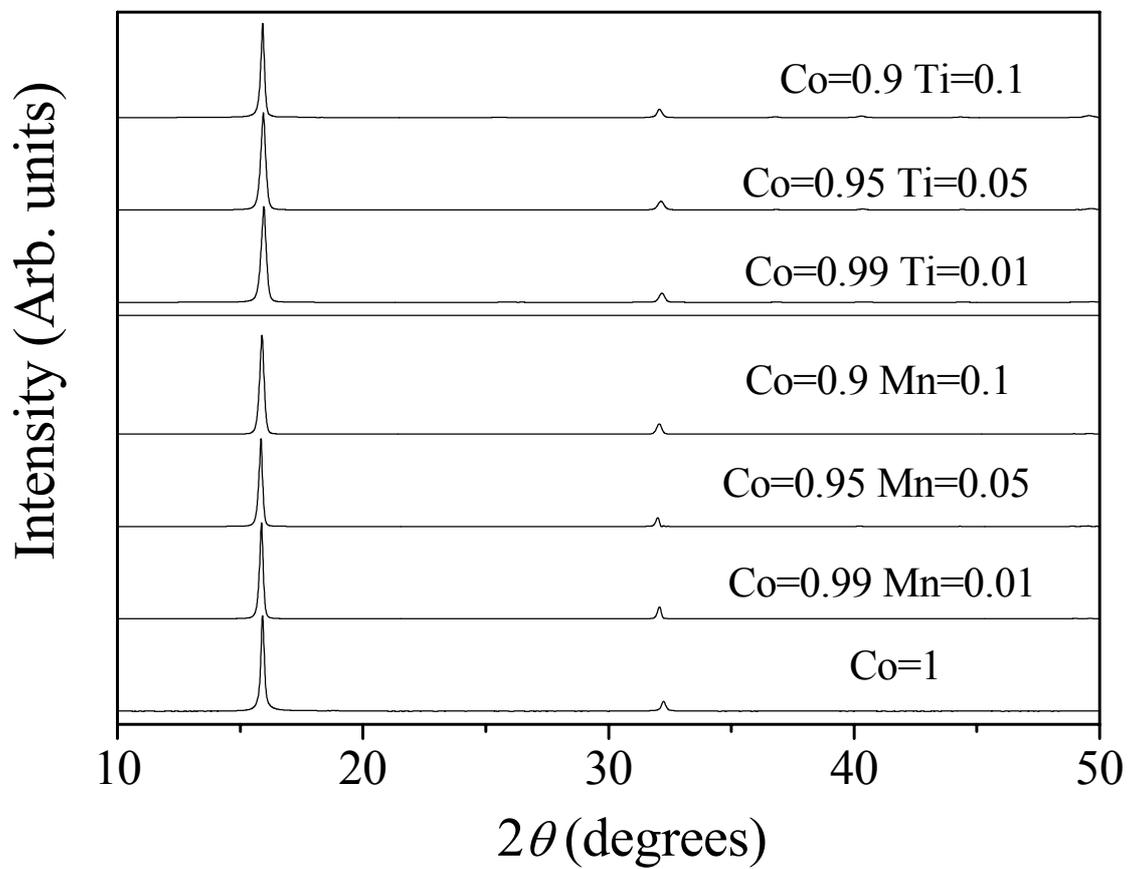

**Figure 4**

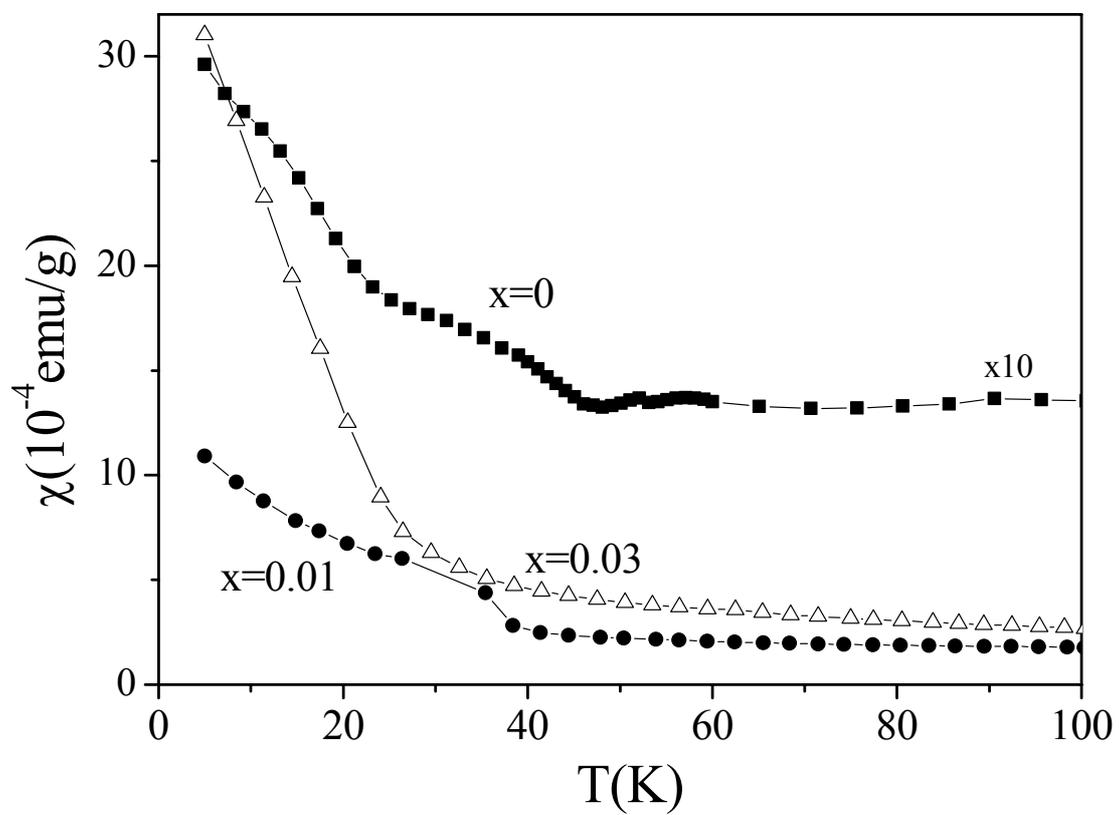

**Figure 5 (a)**

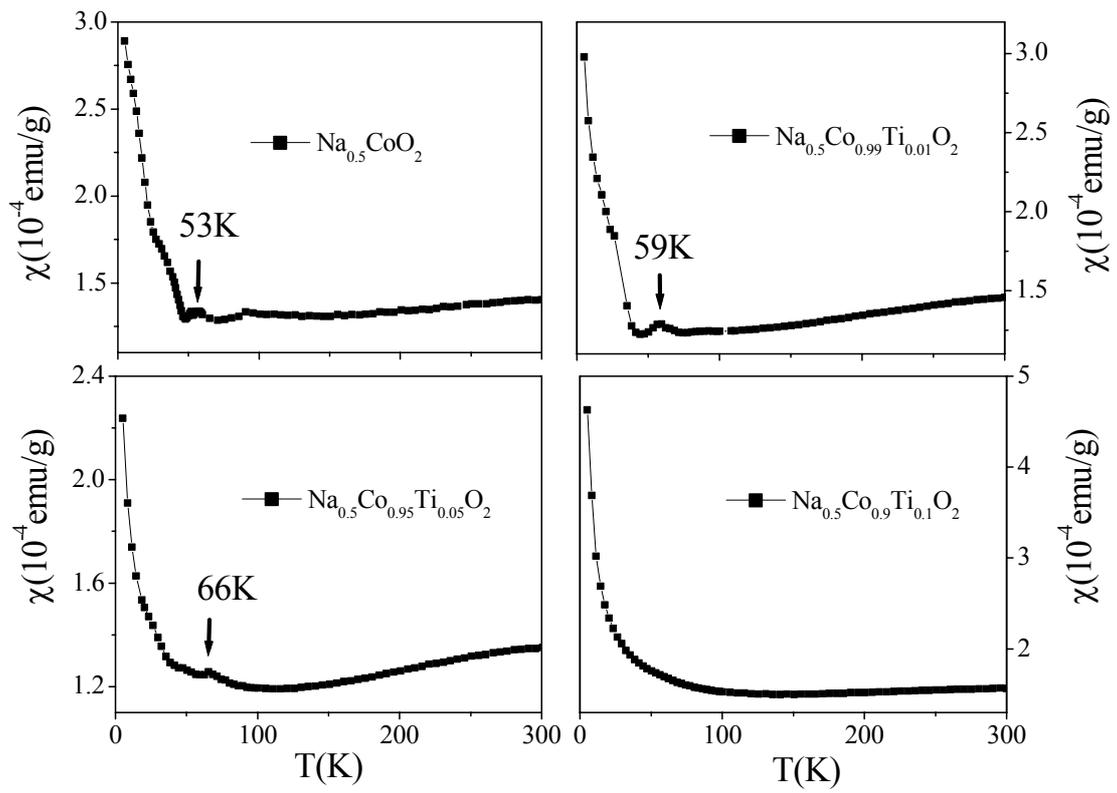

**Figure 5 (b)**

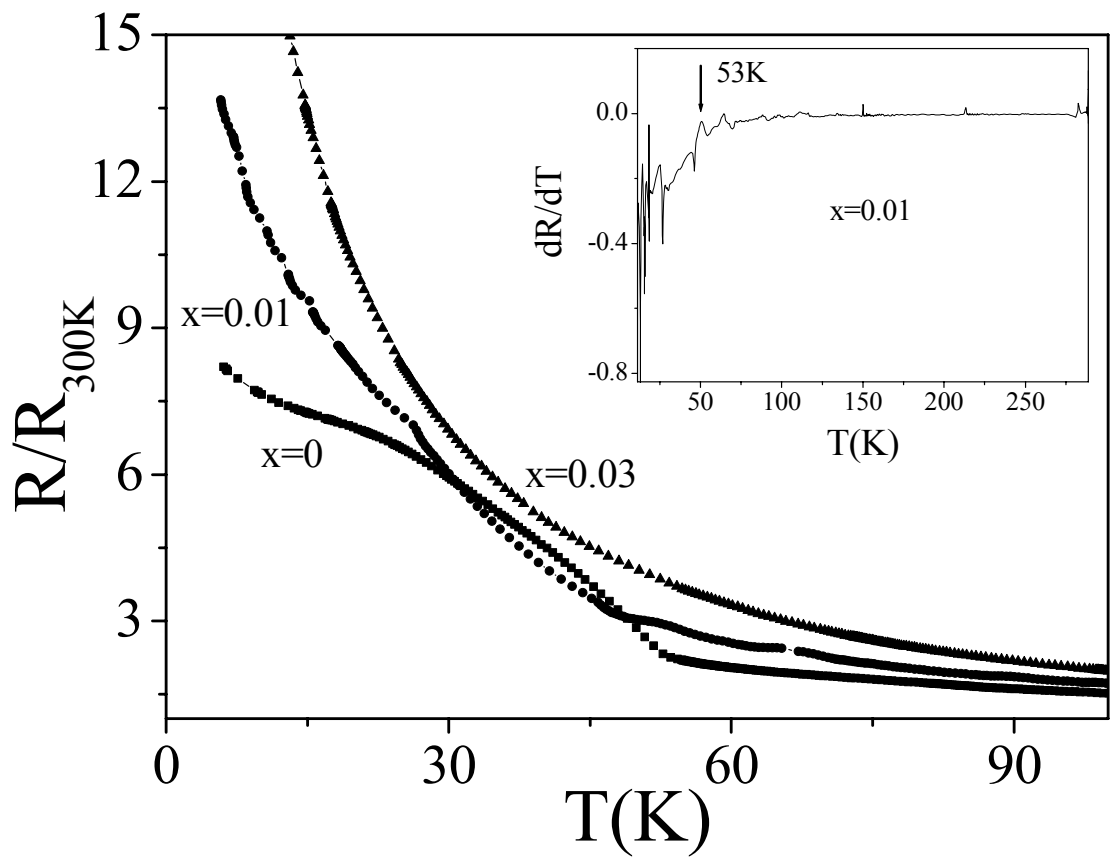

**Figure 6 (a)**

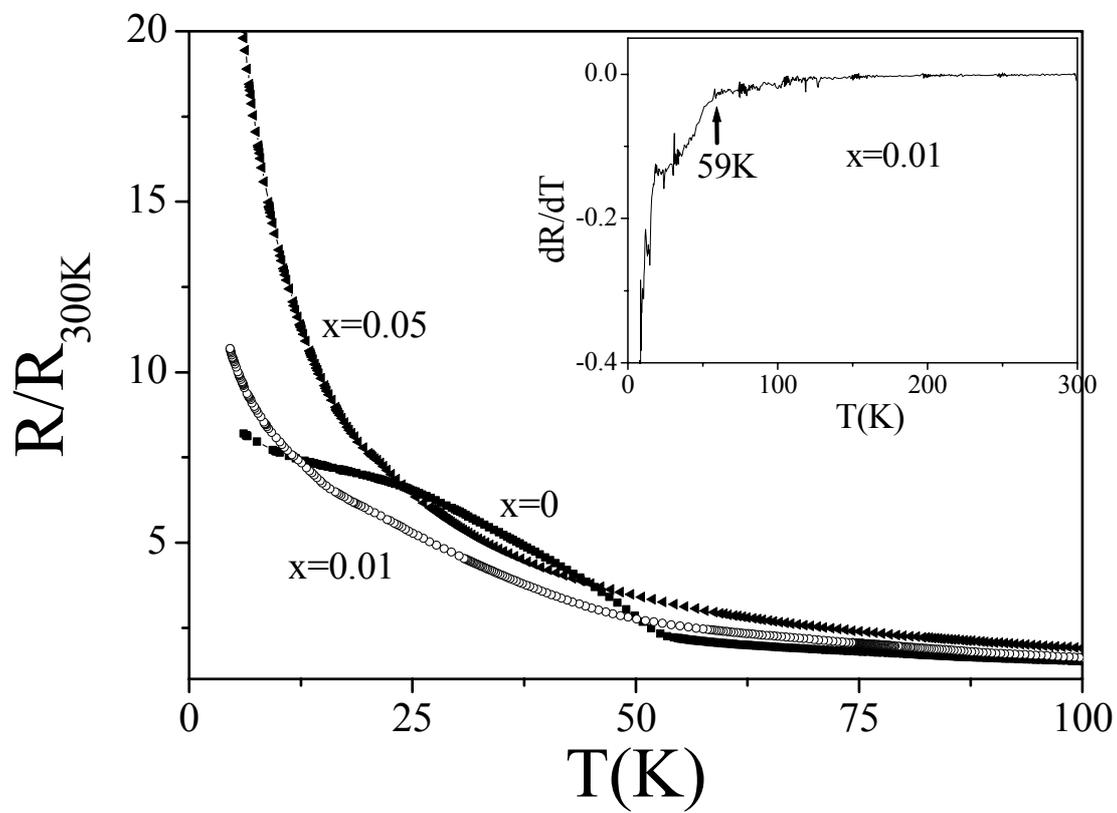

**Figure 6 (b)**